\documentclass[12pt]{article}

\begin{document}                                                     


\def\g{{\tt g}}
\def\1{{\bf 1}}
\def\Z{{\bf Z}}
\def\ee{\end{equation}}
\def\be{\begin{equation}}
\def\l{\label}
\def\D{{\cal D}}
\def\U{{\cal U}}
\def\sin{{\rm sin}}
\def\cos{{\rm cos}}
\def\f{{\bf \Phi}}
\def\v{\varphi}
\def\O{\bf {\cal O}}
\def\C{\bf C}
\def\C{\bf C}
\def\Q{{\cal Q}}
\def\G{{\cal G}}
\def\CP{\bf CP}
\def\e{\rm e}
\def\0{\nonumber}
\def\eea{\end{eqnarray}}
\def\bea{\begin{eqnarray}}
\def\Tr{\rm Tr}
\def\IR{\bf R}
\def\ZZ{\bf Z}
\def\T{\tau}
\def\ep{\epsilon}
\def\k{{\tt k}}




\def\title#1{\centerline{\huge{#1}}}
\def\author#1{\centerline{\large{#1}}}
\def\address#1{\centerline{\it #1}}
\def\ack{{\bf Acknowledgments}}
\def\Bibliography{\begin{thebibliography}}
\def\endbib{\end{thebibliography}}


\begin{titlepage}

\hfill ULB-TH/03-35


\title{On the covariant quantization}
\title{of tensionless bosonic strings}
\title{in AdS spacetime}

\vspace{1.5cm}
\author{Giulio Bonelli}
\vspace{.3cm}
\address{
Physique Theorique et Mathematique - Universite Libre de Bruxelles}
\address{
Campus Plaine C.P. 231; B 1050 Bruxelles, Belgium}
\address{e-mail address: gbonelli@ulb.ac.be}
\begin{abstract}
The covariant quantization of
the tensionless free bosonic (open and closed) strings in AdS spaces is obtained.
This is done by representing the AdS space as an hyperboloid in a flat
auxiliary space and by studying the resulting string constrained hamiltonian 
system in the tensionless limit.
It turns out that the constraint algebra simplifies in the tensionless case in
such a way that the closed BRST quantization can be formulated
and the theory admits then an explicit covariant quantization scheme.
This holds for any value of the dimension of the AdS space.
\end{abstract}
\tableofcontents



\end{titlepage}

\section{Introduction}

The tensionless limit of string theory is as important to be understood as 
the field theory of massless particle excitations is. That's because, as in
field theory the most useful and beautiful symmetries, namely gauge symmetries,
are typical of massless particles field theories, we expect higher
bigger symmetries and nicer quantum properties to appear in the tensionless limit 
of string theory too \cite{G}.

As far as the flat background space is concerned, one finds in fact more than
just new gauge symmetries. In \cite{HT}
the disappearance of the very concept of the critical dimension was noticed
\footnote{Let us recall also the papers \cite{others} where 
a similar, but still less systematic, approach was sketched.}.
That result made clear that the need of fully solving the Liouville theory
to analyze the strings in arbitrary space-time dimension 
\cite{Polyakov} is specific and peculiar of the tensile string. This happens
since in the tensionless limit the conformal anomaly itself gets scaled away.
The link between the tensionless limit of string theory in flat background and 
higher spin field theories was also explored in \cite{FS}.
Moreover, in \cite{iotl}, the arising of a new infinite symmetry has been
explicitly obtained and the tensionless limit of the interacting second
quantized string was analyzed.

In this notes we start studying the problem of the tensionless limit of strings
on the simplest negatively curved spaces, namely AdS spaces.
Let us notice that, since the expansion of the string on curved backgrounds
has been mostly studied in the point-like regime $\alpha'\to0$, we are facing an
almost unexplored subject
\footnote{The approach we take is different to the "Null String"s one by Schild
\cite{Schild} to which the expansion technique developed first in \cite{devega}
applies. The basic difference is that in the "Null String" approach one takes
the tensionless limit at fixed $\sigma$-model variables, while we keep fixed the
string oscillator variables in order to have an a priori control on 
the mass spectrum.}
and the existing expansion methods are therefore useless in the $\alpha'\to\infty$ limit.

Our approach is quite simple and maybe elementary. 
It follows by considering the AdS space an hyperboloid
in an higher dimensional flat auxiliary space. In such a picture we implement
the restriction to the hyperboloid as a lagrangean constraint to the free theory 
in the ambient space.
The natural set-up for the analysis of such a problem turns out then to be
the constrained hamiltonian formalism
\footnote{
The covariant quantization of the bosonic string in flat spacetime
was originally obtained in \cite{KO}.
For a review of string theory in flat space as a constrained
  hamiltonian system, see \cite{marcbook}.}.
It turns out that the constraint algebra structure simplifies in the 
massless/tensionless case
in such a way that the arising of a larger gauge symmetry takes place.
More specifically, it happens that the geometric constraints are second class 
for generic values of the tension parameter, but in the massless/tensionless regime 
it is possible to single out one half of them which, 
together with the reparametrization/Virasoro generators, are first class.
This property opens the way to a well defined BRST covariant quantization of the system
which we develop here.

In few words, the main point is that massless free excitations on AdS can not probe
the strength of the space-time curvature and therefore, in this case, the assignment
of the value of a finite AdS radius should be regarded as a gauge choice
(the assignment of a zero radius being a degenerate gauge fixing condition).
Let us underline again that this property is special of massless/tensionless excitations 
on AdS space-time.

Let us notice that such a phenomenon is linked to the very definition of
the masslessness itself in AdS space (see \cite{FF}) and 
could be rephrased, from the second quantized point of view, by analyzing the
theory in terms of higher spin fields.
In such terms, it should correspond to the phenomenon noticed in \cite{DW}, i.e.
special slope values at which extra gauge 
degrees of freedom arise in higher spin field theories.

Our main result is then a covariant quantization scheme for tensionless strings
in AdS where no restriction to the space-time dimension appears in the form 
of a critical dimension, the constraint algebra being a Lie algebra without any
non trivial central extensions. 

In the following first warm up section, we will treat the easy case
of the scalar free massless particle in AdS in order to clearly explain
the relevant procedure.
In the subsequent section we study the tensionless limit of the free open 
and closed bosonic string in AdS space by extending the method to such a 
more interesting cases.
A final section points out some open questions and possible further developments
for the second quantized tensionless string theory on AdS background
as a theory of interacting higher spin fields and AdS/CFT at null CFT coupling.

\section{Free spinless bosonic massless particles in AdS}

Let us model the dimension $d$ AdS space as an hyperboloid in a flat $d+1$ 
dimensional space. 
Labeling the coordinates in ${\bf R}^{d+1}$ as $x^\mu$, as $\mu=0,\dots,d$, 
the embedding equation is simply 
$$
x^\mu\eta_{\mu\nu}x^\nu=
-x_0^2+x_1^2+\dots+x_{d-1}^2-x_d^2=R^2
$$
which defines the quadratic form $\eta$. We will usually write $x^2$ for
$x^\mu\eta_{\mu\nu}x^\nu$ and $uv=u^\mu\eta_{\mu\nu}v^\nu$ for various (co)vectors.

The system of a spinless bosonic massless particles in AdS
is most symmetrically described in the hamiltonian formalism.
Actually we can study it as a constrained hamiltonian system of a 
free massless bosonic particle in ${\bf R}^{d+1}$
constrained to the AdS hyperboloid of dimension $d$.
The natural set of constraints (which follow from the Dirac consistency 
procedure applied to the massive free particle with lagrangean constraint
$x^2=R^2$ and by the subsequent zero mass limit and the removal of the Lagrange
multiplier\footnote{To be rigorous, the removal of the Lagrange multiplier should be performed at
quantum mechanical level. For our porpuses, anyway, it is still consistent to perform
it already at classical level.}) are 
\be
p^2=0,\quad x^2-R^2=0\quad{\rm and}\quad  xp=0
\ee
which, together with a further non-singular gauge choice -- for example a light cone gauge 
$vx=\tau$, with $\tau$ the Hamiltonian time coordinate --
give a good hamiltonian formulation of the system, i.e. form a
non degenerate set of constraints.
Differently from the massive case, we can single out from the above set a
smaller one of first class constraints, namely $p^2=0$ and $xp=0$
whose Poisson bracket algebra is closed.
This shows that the spinless massless free particle on AdS is aware of being on
a constant negatively curved space, but is not able to feel the strength of such a
curvature.
Put in another way, the assignment of the AdS radius can be viewed as a gauge fixing
condition for the system.

As far as the quantization is concerned, 
we promote the canonical pair $(x^\mu,p_\mu)$ to operators obeying
the Heisenberg algebra $[x^\mu,p_\nu]=i\delta^\mu_\nu$
as well as the constraint functions to the hermitian operators
$p^2$ and $\frac{1}{2}(xp+px)$.
Notice that the first class constraint algebra still closes, 
namely\footnote{Notice that if a mass term would be added
as $p^2\to p^2+m^2$ then such a property would not be true anymore
and a more elaborate scheme, from the point of view of the constrained 
system analysis, should be applied.}
$[p^2,\frac{1}{2}(xp+px)]=-i2p^2$.
This enables us to perform the Dirac quantization of the dynamical system, that
is to formulate the consistent wave equation system
\be
p^2\psi=0 \quad {\rm and} \quad \left[\frac{1}{2}(xp+px)+\k\right]\psi=0,
\l{Dirac}\ee
where $\k$ is a c-number.
Choosing hyperbolic polar coordinates in ${\bf R}^{d+1}$
one can solve the radial dependence of the wave function 
by the second of Eq.s(\ref{Dirac}) and stay with a
single resulting wave equations in AdS intrinsic coordinates.

There are in fact several possible
generalizations of the above constraint algebra (for example one can add 
the coupling with a constant curvature bulk electromagnetic field)
and these could be interesting constrained dynamical systems to study
as well.

It is also possible rephrase the quantization procedure at a BRST level.
This is done by 
introducing the independent anticommuting ghost real pairs $(c,b)$ and $(c',b')$ 
and encoding the constraint algebra in the BRST charge
$$
\Q=\frac{1}{2}cp^2 + c'\left[\frac{1}{2}(xp+px)-i(cb-bc)+k\right]
\quad {\rm obeying} \quad \Q^2=0
$$
where the antisymmetric ordering for the real ghosts has been explicitly
applied, $k$ is a real number and the pure ghost part of the BRST
charge has been chosen to be minimal.
The anti-hermitian ghost number operator is
$
\G=\frac{1}{2}[c,b]+\frac{1}{2}[c',b']
$
and is such that $[\G,\Q]=\Q$.
The BRST state cohomology $Ker\Q/Im\Q$ can be easily calculated 
to reproduce\footnote{Assuming standard partial invertibility 
for the relevant operators which generalizes the usual ones holding
for the free massless relativistic particle in flat space.} eq.(\ref{Dirac})
with $\k=k+i$ on the lowest ghost number state.

\section{Tensionless bosonic free strings in AdS}

After the previous warm up section, let us now enter the main subject of the
present note, namely the tensionless limit of free bosonic strings in AdS space.
We will follow the same strategy developed in the previous section for the
spinless scalar particle and generalize it to open and closed free bosonic strings.

\subsection{Open strings}

\subsubsection{The constraint algebra}

In this subsection 
we analyze the constrained system of open strings in ${\bf R}^{d+1}$
bound to stay on the hyperboloid $x^2=R^2$ in the tensionless limit.
The study of this system amounts to the extraction of the leading order 
terms in $\alpha'$ out of the Virasoro constraints
\be
\frac{1}{2}\left[2\pi\alpha'P^2+\frac{1}{2\pi\alpha'}(X')^2\right]
\quad {\rm and}\quad 
X'P
\l{virasoro}\ee
and the geometric constraints\footnote{These can be shown to follow from the
Dirac consistency procedure applied to the string lagrangian in 
${\bf R}^{d+1}$ augmented by the lagrangian constraint $X^2-R^2=0$. The
Lagrange multiplier is already fixed at classical level.}
\be
\frac{1}{\alpha'}\left(X^2-R^2\right)
\quad {\rm and}\quad 
XP
\l{geometric}\ee
Then, following the method explained in the previous section, we have to
select out of the geometric constraints a subset such that, together with 
the contracted Virasoro constraints, one gets a closed algebra under commutation
so that
one is left with first class constraints only and a BRST quantization
procedure is available. 
Let us notice that such a procedure was available for massless particles 
only on AdS and is respectively available for the tensionless strings only in
AdS (while it is not for the tensile case when the subleading terms have to be kept).
Moreover, since first class constraints weight twice 
-- in the degrees of freedom counting --
than second class ones (see for a review \cite{chs}), 
to get the correct degrees of freedom counting, the subset of 
constraints that we have to single out has to contain one half of the original
geometric ones.

Before proceeding, let us now just fix some notation.
The system is given by the string center of mass
variables $(x_\mu,p^\mu)$ and the infinite set of oscillators 
$(a_{n\mu},a^*_{n\mu})$, with $n>0$.
They satisfy the usual canonical commutation relations (CCRs)
\be
[x_\mu,p^\nu]=i\delta_\mu^\nu
\quad , \quad
[a_{n\mu},a^*_{m\nu}]=\eta_{\mu\nu}\delta_{nm}
\l{ccr}\ee
and the other commutators are vanishing.
The string coordinate and its conjugate momentum are ($\sigma\in [0,\pi]$)
$$
X_\mu(\sigma)=x_\mu +\sum_{n>0} \sqrt{\frac{2\alpha'}{n}}\left(a_{n\mu}+a_{n\mu}^*\right)cos(n\sigma),
$$ 
$$
P^\mu(\sigma)=\frac{p^\mu}{\pi} +
\frac{1}{i\pi}\sum_{n>0} \sqrt{\frac{n}{2\alpha'}}\left(a_{n\mu}-a_{n\mu}^*\right)cos(n\sigma).
$$
Let us notice that we define the string expansion in the flat auxiliary space
as the free string expansion. This is the only correct possibility once we
required that 
the tension parameter enters the oscillator expansion of the string
canonical variables 
independently on the further constraints fixing the curved space the string 
will be constrained on, that is independently on the geometric constraints.

The extraction of the leading order in the large $\alpha'$ expansion
of the Virasoro constraints (\ref{virasoro}) is exactly
equal to the calculation in the flat space (see \cite{HT,iotl}), i.e. we have (after rescaling)
$$ L_0=p^2, \quad 
L_n=p\cdot a_n
\quad {\rm and}\quad 
L_n^*=p\cdot a_n^*.$$

A choice of one half constraints out of the geometrical ones
(\ref{geometric}) is to select the
zeromode part of $XP$ and the oscillating modes of $\frac{1}{\alpha'}\left(X^2-R^2\right)$, that is 
$$
D_0=\int_0^\pi d\sigma :XP:(\sigma)=\frac{1}{2}(xp+px)+\frac{1}{2i}
\sum_{n>0}\left(a_n^2-{a^*}_n^2\right)
$$
and the leading order coefficients in the Fourier expansion of 
$\frac{1}{\alpha'}:XX':(\sigma)=\sum_{p>0}\sin(p\sigma)\left(D_p+o(\frac{1}{\sqrt{\alpha'}})\right)$, namely
$$
D_p=
\sum_{n>0}\left(\sqrt{\frac{p+n}{n}}-\sqrt{\frac{n}{p+n}}\right)
\left(a_{p+n}a_n+a^*_{p+n}a_n+a^*_na_{p+n}+a^*_{p+n}a^*_n\right)
+$$ $$+
\sum_{m=1}^{p-1}\sqrt{\frac{m}{p-m}}
\left(a_ma_{p-m}+a^*_{p-m}a_m+a^*_ma_{p-m}+a^*_{p-m}a^*_m\right)$$
where $p$ runs over the positive integers (and the second sum is not there for $p=1$).
Notice that $D_0$ and $D_p$ are hermitian.
The left over constraints, i.e. the leading terms in the zero mode part of
$\frac{1}{\alpha'}\left(X^2-R^2\right)$ 
and in the oscillatory part of $XP$, can therefore be regarded as gauge fixing conditions.

To prove that the above choice of constraints is first class, we have to
exhibit the closure of the constraint algebra generated by the $L$s and
the $D$s. The algebra can be checked to be given by
\be
[L_0,D_0]=-2iL_0 ,\quad
[L_0,D_p]=0 ,\quad
[L_n,D_0]=-i(L_n-L^*_n),
\l{ext1}\ee
\be
[L_n,D_p]=\left(\sqrt{\frac{p+n}{n}}-\sqrt{\frac{n}{p+n}}\right)
(L_{p+n}+L^*_{p+n})+\zeta_{n,p}((L_{|p-n|}+L^*_{|p-n|})
\l{ext2}\ee
where $\zeta_{n,p}=\frac{p}{\sqrt{n|n-p|}}$ if $p\not=n$ and $\zeta_{p,p}=0$.
Moreover\footnote{To obtain the following commutation relations the reader might start
from the commutation relations 
$$
\left[XX'(\sigma),XX'(\sigma')\right]=0 \quad {\rm and}
\quad
\left[\int PX , XX'(\sigma)\right]=-2iXX'(\sigma),
$$ 
pass to the normal ordered expressions
and then consider the leading order terms.} we have
\be
[D_0,D_p]=-2i\left(D_p-\frac{d+1}{2}\left(1+(-1)^p\right)\right)
\quad {\rm and}\quad
[D_p,D_q]=0.
\l{internal}\ee
Notice that quantum mechanically one has to consider the operator $D_p$
to be defined up to an additional constant due to normal ordering ambiguity.
This can be added to cancel the c-number term appearing in the first
commutator in (\ref{internal}). We will fix the actual value of such a
quantity together with the proper ghost contribution by building a nilpotent
quantum BRST charge for the system under consideration.

\subsubsection{The quantum BRST charge}

In order to build the BRST charge of the system,
we introduce the relative anticommuting ghosts $c_n$, $c_n^*$ and $c_0$ 
and the anti-ghosts $b_n$, $b^*_n$ and $b_0$ (normalized by $[c_0,b_0]_+=1$,
$[c_m,b^*_n]_+=\delta_{mn}$, $[c^*_n,b_m]_+=\delta_{nm}$ and other anti-commutators vanishing)
for the contracted Virasoro constraints $L_n$, $L^*_n$ and $L_0$
as well as the hermitian ghosts $c'_p$ and $c'_0$ and the relative hermitian
anti-ghosts $b'_p$ and $b'_0$ (normalized by $[c'_0,b'_0]_+=1$,
$[c'_n,b'_m]_+=\delta_{nm}$ and other anti-commutators vanishing)
for the contracted geometrical embedding constraints $D_p$ and $D_0$.
The ordering prescription that we keep for hermitian ghosts is the usual anti-symmetrization.

As it results from \cite{HT,iotl}, the BRST charge corresponding to the 
tensionless open string constraints in $R^{d+1}$ is given by
$$
\Q_{open,R^{d+1}}=\frac{1}{2}c_0L_0+
\sum_{n>0}\left[L_n^*c_n+c_n^*L_n-2c^*_nc_nb_0\right]
$$
and keeps into account the contracted Virasoro constraints only.
The full BRST charge implementing the above full constrained
system is obtained then by adding the geometric constraints
as an improvement. The way we find natural to proceed is to notice that, since 
the algebra (\ref{ext1}-\ref{ext2}) is of the form $[L,D]=L$, then
the geometric constraints $D$'s can be promoted to operators commuting 
with $\Q_{open,R^{d+1}}$ by adding suitable (bc) ghost terms.

Let us therefore introduce the following invariant bilinear combinations
\be
l_{(mn)}=a_m\cdot a_n -c_mb_n-c_nb_m
\quad
l^*_{(mn)}=a^*_m\cdot a^*_n +c^*_mb^*_n+c^*_nb^*_m
\l{rota}\ee
$$
h_{mn}=a^*_m\cdot a_n+c^*_mb_n+b^*_mc_n+\frac{d-1}{2}\delta_{mn}
$$
and\footnote{
The above family of quadratic combinations close to form the following $sp(\infty)$ algebra
\be
[l_{(mn)}, l_{(pq)}]=0 \quad
[l_{(mn)}, h_{pq}]=\delta_{np}l_{(mq)}+\delta_{mp}l_{(nq)}
\l{symal}\ee
$$
[h_{mn}, h_{pq}]=\delta_{np}h_{mq}-\delta_{mq}h_{pn}
$$
$$
[l^*_{(mn)}, l^*_{(pq)}]=0 \quad
[l^*_{(mn)}, h_{pq}]=-\delta_{nq}l^*_{(mp)}-\delta_{mq}l^*_{(np)}
$$
$$
[l_{(mn)},l^*_{(pq)}]= h_{qm}\delta_{np}+
h_{pm}\delta_{nq}+h_{qn}\delta_{mp}+h_{pn}\delta_{mq}
$$
which is useful to check the algebraic calculations.
Notice that all these bilinears commute with the unconstrained BRST
charge $\Q_{open,R^{d+1}}$.}
let us define the ghost completition of the contracted geometric constraints 
as\footnote{As in the easier case of the scalar particle, we can add a real
constant to the definition of $\Delta_0$.}
$$
\Delta_0=\frac{1}{2}\left(xp+px\right)-i[c_0,b_0]+i\sum_{m>0}
\left(b^*_mc_m-c^*_mb_m\right)+\frac{1}{2i}\sum_{n>0}\left(l_{(nn)}-l^*_{(nn)}\right)
-\frac{3}{2}i
$$
and 
$$
\Delta_p=\sum_{n=1}^{p-1}\sqrt{\frac{p-n}{n}}
\left(l_{(n,p-n)}+l^*_{(n,p-n)}+h_{n,p-n}+h_{p-n,n}\right)
+ $$ $$ +
\sum_{n>0}\left(\sqrt{\frac{p+n}{n}}-\sqrt{\frac{n}{p+n}}\right)
\left(l_{(n,p+n)}+l^*_{(n,p+n)}+h_{n,p+n}+h_{p+n,n}\right)
$$
The commutation algebra satisfied by the above operators is
\be
[\Delta_0,\Delta_p]=-2i\Delta_p
\quad{\rm and}\quad
[\Delta_p,\Delta_q]=0
\l{int2}\ee
We have now a clear synthetic framework to obtain the full quantum BRST charge
$$
\Q_{open,AdS}=\Q_{open,R^{d+1}}+c'_0\Delta_0 +\sum_{p>0}c'_p\Delta_p
+ic'_0\sum_{p>0}[c'_p,b'_p]
$$
Notice that since the ghost completed geometrical constraints $\Delta$s commute with the
unconstrained BRST charge $\Q_{open,R^{d+1}}$ and fulfill the algebra
(\ref{int2}) we have $$\Q_{open,AdS}^2=0$$ 
irrespectively to the value of the space-time dimension.
This shows that, as in the flat case, also on AdS space 
the whole conformal anomaly scales away and the problem of
critical dimension does not exist anymore in the tensionless limit.

Moreover, since the expressions we started from were invariant under the
AdS rotational group $SO(2,d-1)$, also the tensionless limit is.
More concretely, one can check that the $SO(2,d-1)$ generators
\be 
J_{\mu\nu}=\frac{1}{2}(x_\mu p_\nu -x_\nu p_\mu)-\frac{i}{2}\sum_{n>0}
(a^*_{n\mu}a_{n\nu}-a^*_{n\nu}a_{n\mu})
\l{rotgen}\ee
are such that $[J_{\mu\nu},\Q_{open,AdS}]=0$.

\subsection{Closed strings}

In order to treat the closed string case, we follow a procedure analogous
to the one we developed in the open string case.
Let us expand the $\sigma$-model canonical coordinates in oscillators,
that is (now $\sigma\in[0,2\pi]$)
$$
X^\mu(\sigma)=x^\mu+\sum_{n>0}\sqrt{\frac{\alpha'}{2n}}
\left(
a_n^\mu e^{-in\sigma} +\bar a_n^\mu e^{in\sigma} 
+\quad {\rm h.c.}\right)
$$
$$
P^\mu(\sigma)=\frac{p^\mu}{2\pi}+\frac{1}{2\pi}
\sum_{n>0}\sqrt{\frac{n}{2\alpha'}}
\left(
-ia_n^\mu e^{-in\sigma} -i\bar a_n^\mu e^{in\sigma} 
+
\quad {\rm h.c.}\right)
$$
where the above modes satisfy the CCRs
\be
[x^\mu,p^\nu]=i\eta^{\mu\nu}
\quad
[a_n^\mu,(a^*)^\nu_m]=\eta^{\mu\nu}\delta_{nm}
\quad
[\bar a_n^\mu,(\bar a^*)^\nu_m]=\eta^{\mu\nu}\delta_{nm}
\l{cccr}\ee
and the other commutators vanishing.

Calculating the Virasoro constraints and performing the leading order terms
extraction as we did in the open string case, we find that 
the left over constraints for the tensionless closed string are 
$$ p^2, \quad L_n=p\cdot a_n, \quad  L_n^*=p\cdot a_n^*, $$
$$ \bar L_n=p\cdot\bar a_n, \quad  \bar L_n^*=p\cdot\bar a_n^*, 
\quad N-\bar N=\sum_{n>0} n \left(a^*_n\cdot a_n
-\bar a^*_n\cdot \bar a_n\right),$$
the last one being the usual level matching condition (and we don't define any
$L_0=\bar L_0=p^2$, but we just keep denoting $p^2$ not to cause extra
confusion with the level matching and the notation which makes sense within
the tensile string usual notation).

As far as the geometric constraints
$$
\frac{1}{\alpha'}\left(X^2-R^2\right)
\quad {\rm and}\quad 
XP
$$
are concerned, one can work out easily their leading order factors in the
tensionless limit and, as in the open free string case, single out
the oscillatory part of the first and the zero mode part of the second, that
is the leading order of $\frac{1}{\alpha'}:XX':(\sigma)$ and $\oint :XP:$.

Calculating the relevant Fourier modes and leading terms, 
we have $\frac{1}{\alpha'}:XX':(\sigma)=
\frac{i}{2}\sum_{p>0}\left(C_pe^{ip\sigma}-C_p^*e^{-ip\sigma}\right)
+o\left(\frac{1}{\sqrt{\alpha'}}\right)$,
where
$$
C_p=\sum_{m=1}^{p-1}\sqrt{\frac{m}{p-m}}
\left(\bar a_{p-m}+a^*_{p-m}\right)\cdot
\left(\bar a_{m}+a^*_{m}\right)+
$$
\be
+\sum_{n>0}\left(\sqrt{\frac{p+n}{n}}-\sqrt{\frac{n}{p+n}}\right)
\left(a_{n}+\bar a^*_{n}\right)
\cdot
\left(\bar a_{p+n}+a^*_{p+n}\right)
\l{clco}\ee
and
$$
C_0=\oint :XP:=
\frac{1}{2}\left(xp+px\right)
+i\sum_{n>0}
\left(\bar a_n^* a_n^*-\bar a_{n} a_{n}\right)
$$
Notice that $C_0$ is hermitian, while $C_p$ is conjugated to $C_p^*$ and viceversa.
(We don't explicitly write the normal ordered expression for $C_p$ since it
coincides with the one we gave above).

It is straightforward to verify that the algebra of the constraint functions
closes to a Lie algebra and that therefore the BRST quantization procedure
can be developed as in the open string case.
This shows that also in the closed string case it is possible to single out a subset 
of the geometrical constraints in order to present the constrained hamiltonian system 
just in terms of first class constraints only.
Specifically, the actual form of the constraint algebra is given by the following commutation relations
$$
[L_n,L^*_m]=p^2\delta_{nm},
\quad
[\bar L_n,\bar L^*_m]=p^2\delta_{nm},
$$
$$
[L_n,N-\bar N]=nL_n ,
\quad
[\bar L_n,N-\bar N]=-n\bar L_n ,
$$
$$
[p^2,C_0]=-2ip^2 ,
\quad
[L_n,C_0]=-iL_n+i\bar L^*_n ,
\quad
[\bar L_n,C_0]=-i\bar L_n +iL_n^* ,
$$
$$
[L_n,C_p]=\left(\sqrt{\frac{n}{n-p}}-\sqrt{\frac{n-p}{n}}\right)
\left(L_{n-p}+\bar L^*_{n-p}\right) \quad{\rm if}\quad n>p
$$ 
$$
[L_n,C_p]=\left(\sqrt{\frac{p-n}{n}}+\sqrt{\frac{n}{p-n}}\right)
\left(\bar L_{p-n}+ L^*_{p-n}\right) \quad{\rm if}\quad n<p
$$
$$
[L_p,C_p]=0\quad [\bar L^*_n,C_p]=[L_n,C_p]
$$
$$
[\bar L_n,C_p]=[L_n^*,C_p]=\left(\sqrt{\frac{p+n}{n}}-\sqrt{\frac{n}{p+n}}\right)
\left(\bar L_{n+p}+L^*_{n+p}\right)
$$
$$
[C_0,C_p]=-2iC_p
$$
the others being vanishing or can be obtained by hermitian conjugation of the above ones.
Notice that the above algebra is not a direct product of two copies of the
open one. Here, in fact, we see explicitly how the curvature of the background
affects the left/right sectors mixing (as it is expected to happen from a more
general point of view).

Also the construction of the quantum BRST charge follows a path similar to the
open string case.  
Introducing the ghosts for the
geometric constraints $c'_p$, $c'_0$ and $(c')^*_p$ as well as the
conjugated anti-ghosts $(b')^*_p$, $b'_0$ and $b'_p$,
the BRST charge can be built as
\be
\Q_{closed,AdS}=\Q_{closed,R^{d+1}}+\sum_{p>0}\left((c')^*_p\Gamma_p
+\Gamma^*_pc'_p\right) + c'_0\Gamma_0 + [[b'c'c']]
\l{clobrst}\ee
where
$$
\Q_{closed,R^{d+1}}=\frac{1}{2}c_0p^2+\sum_{n>0}
\left(c^*_n pa_n+ pa_n^*c_n +\bar c^*_n p\bar a_n+ p\bar a_n^*\bar c_n \right)
+$$ $$
+\hat c_0\sum_{n>0}n\left(a^*_na_n+c^*_nb_n+b^*_nc_n-
\bar a^*_n\bar a_n-\bar c^*_n\bar b_n-\bar b^*_n\bar c_n\right)
-2b_0\sum_{n>0}n\left(c^*_nc_n+\bar c^*_n\bar c_n\right),
$$
$\Gamma_p=C_p+[{\rm ghosts}]$ are the ghost completition of the constraints
$C_p$ such that $[\Q_{closed,R^{d+1}},\Gamma_p]=0$ and still satisfy the internal 
algebra of the constraints, namely
\be
[\Gamma_0,\Gamma_p]=-2i\Gamma_p,\quad
[\Gamma_0,\Gamma^*_p]=-2i\Gamma^*_p
\l{intaug}\ee
and the others vanishing.
Finally, the term $[[b'c'c']]$ is built from the algebra (\ref{intaug}) structure constants
and is equal to $-2ic'_0\sum_{p>0} \left((b')^*_pc'_p-(c')^*_pb'_p\right)$.

Since the algebra (\ref{intaug}) is fulfilled and the $\Gamma$s commute with 
$\Q_{closed,R^{d+1}}$, then the BRST charge (\ref{clobrst}) satisfies
$$\Q_{closed,AdS}^2=0$$
identically.

Notice moreover that all the constraints commute with the generators of the AdS
isometry group $SO(2,d-1)$, i.e. with
$$
J_{\mu\nu}=\frac{1}{2}\left(p_\mu x_\nu-p_\nu x_\mu\right)
+\frac{i}{2}\sum_n
\left(a^*_{n\mu}a_{n\nu}-a^*_{n\nu}a_{n\mu}
+\bar a^*_{n\mu}\bar a_{n\nu}-\bar a^*_{n\nu}\bar a_{n\mu}\right)
$$
which therefore commute with the BRST charge too.

\section{Conclusions and Open Questions}

In this short note we have studied few basic properties of tensionless bosonic
strings in AdS space-time, namely the very existence of a covariant quantization
scheme for such a system. We found that
the special simplification in the constraint algebra that takes place in such a limit
enables an explicitly covariant quantization scheme of the theory in any 
spacetime dimension.
Let us notice that the light-cone gauge fixed approach developed in
\cite{CKKY}, although seemingly not rigorous, indirectly suggests 
such a result
\footnote{Notice that a different
approach to the tensionless string in AdS
appeared also in \cite{LZ}.
That approach is a limiting case of the construction carried out in \cite{FL}
and is based on a coset WZNW realization of the AdS $\sigma$-model
which therefore implies the presence of a specific balancing background
antisymmetric field. It follows that such a approach refers to a different
set up with respect to the one we have been considering in this paper.}.

In principle, the results that we obtained can be extended to any spacetime 
which can be obtained as a quadric in a higher dimensional flat space, as
dS space for example.
The physical difference between the positive and negatively curved space
have to be understood from the explicit calculation of the spectrum of the theories,
that is from the calculation of the BRST state cohomology whose structure
naturally depends on the signature of the defining quadratic form $\eta$.

The second quantization of the free tensionless string theories that we just developed
results in an infinite chain of higher spin theories on AdS 
(about this subject see \cite{V}\cite{HS}\cite{BMV}\cite{md}\cite{mm}).
In such a general framework it is possible indeed to study 
also the interaction of tensionless strings as already proposed in \cite{iotl}. 
Actually, we can extend at some rate the arguments about string fragmentation
and world-sheet picture instability by \cite{Gross2} which indicate that the first
quantized interacting string gets literally undone in the tensionless limit.
This phenomenon has a clear analogous in QED which is the IR-catastrophe
in first quantization when finite energy amounts can be emitted in the form of an 
infinity of soft photons. Notice that some caveat has to be raised here, since 
perturbative IR fluctuations are typically dumped in negatively curved
space-times \cite{CW} and therefore the analogous of the analysis in \cite{Gross2}
in AdS space should be carried out carefully possibly giving a tendential
dumping of the string fragmentation effect.
Such an unclear picture promotes the string field approach to be the natural 
framework to correctly formulate tensionless strings interactions.
In order to do this, one has to extend the usual string field theory techniques to AdS 
to build the three string vertex \cite{Witten,GJ}.
The value of the AdS radius is expected to play a role in the interacting
theory due to the deformation of the gauge symmetry.

The approach we developed could in fact be extended to superstring theories
too with results naturally much similar to the ones we have found for the
bosonic case. In particular one could consider the tensionless limit of 
type IIB string on $AdS_5\times S^5$ (which can be easily described as a product of
two quadrics in a twelve dimensional space) in order to get some new inputs 
for AdS/CFT in the tensionless limit\cite{holo}\cite{adscft}\cite{also}.
The second quantization of the tensionless string theories that we 
obtained here could in fact shed some light on that subject.
Actually, to advance in such a direction one should study the spectrum 
of the tensionless theory, that is (from our perspective)
one has to solve the state cohomology of the BRST charge $\Q_{open,AdS}$ or
$\Q_{closed,AdS}$, and compare it with the spectrum of conserved currents
of a proper boundary theory.
The actual calculation of the BRST state cohomology is obviously the first
development to be carried out. It is clear that this will classify the
physical states in terms of IURs of the AdS group with a spectrum which we
expect highly symmetric and irreducible to a deformation of the flat space
string spectrum, as the results in \cite{BMV} already suggest.

These and other possible aspects of the tensionless limit of string theories
are open issues for further research.

\vspace{.5cm}
\noindent
\ack

\noindent
I would like to thank G.~Barnich, X.~Bekaert, M.~Bertolini, N.~Boulanger,
S.~Cnockaert, M.~Grigoriev, A.~Hammou, M.~Henneaux, L.~Houart, A.~Sagnotti, 
D.~Sorokin and M.~Tsulaia 
for stimulating discussions and encouragement. In particular I would like to
thank M.~Henneaux for a careful reading of the manuscript and useful remarks
and A.~Hammou for enlighting discussions.
\noindent
This work is supported by the Marie Curie fellowship contract
HPMF-CT-2002-0185.
Work supported in part by the ``Actions de Recherche
Concert{\'e}es" of the ``Direction de la Recherche Scientifique -
Communaut{\'e} Fran{\c c}aise de Belgique", by a ``P\^ole
d'Attraction Interuniversitaire" (Belgium), by IISN-Belgium
(convention 4.4505.86)  and by the European Commission RTN programme
HPRN-CT-00131, in which G.B. is associated to K. U. Leuven.


\small

\Bibliography{99}

\bibitem{G}
D.~J.~Gross,
``High-Energy Symmetries Of String Theory,''
Phys.\ Rev.\ Lett.\  {\bf 60} (1988) 1229.

\bibitem{HT} 
M.~Henneaux and C.~Teitelboim,
``First And Second Quantized Point Particles Of Any Spin,''
In "Santiago 1987, Proceedings, Quantum mechanics of fundamental systems 2", pp. 113-152. 
Edited by C. Teitelboim and J. Zanelli, Plenum Press.

\bibitem{others} 
S.~Ouvry and J.~Stern
Phys.\ Lett.\ {\bf B} 177 (1986) 335;
A.~K.~H.~Bengtsson,
Phys.\ Lett.\ {\bf B} 182 (1986) 321.

\bibitem{Polyakov}
A.~M.~Polyakov,
``Quantum Geometry Of Bosonic Strings,''
Phys.\ Lett.\ B {\bf 103} (1981) 207.

\bibitem{FS}
D.~Francia and A.~Sagnotti,
``On the geometry of higher-spin gauge fields,''
Class.\ Quant.\ Grav.\  {\bf 20} (2003) S473
[arXiv:hep-th/0212185].

\bibitem{iotl} G.~Bonelli,
``On the tensionless limit of bosonic strings, infinite symmetries and  higher spins,''
Nucl.\ Phys.\ {\bf B} 669 (2003) 159.
[arXiv:hep-th/0305155].

\bibitem{Schild}
A.~Schild,
``Classical Null Strings,''
Phys.\ Rev.\ D {\bf 16} (1977) 1722.

\bibitem{devega}
H.~J.~de Vega and N.~Sanchez,
Phys.\ Lett.\ B {\bf 197} (1987) 320.

\bibitem{KO}
M.~Kato and K.~Ogawa,
``Covariant Quantization Of String Based On BRS Invariance,''
Nucl.\ Phys.\ B {\bf 212} (1983) 443.

\bibitem{marcbook}
L.~Brink and M.~Henneaux,
``Principles of String Theory'' (part II),
Series of the Centro de Estudios Cientificos de Santiago,
Plenum Press (1988). 

\bibitem{FF}
S.~Ferrara and C.~Fronsdal,
``Conformal fields in higher dimensions,''
arXiv:hep-th/0006009.

\bibitem{DW}
S.~Deser and A.~Waldron,
Phys.\ Rev.\ Lett.\  {\bf 87} (2001) 031601
[arXiv:hep-th/0102166].

\bibitem{chs} A.~ Hanson, T.~Regge and C.~Teitelboim,
``Constrained Hamiltonian Systems''
Contributi del Centro Linceo Interdisciplinare di Scienze Matematiche e loro 
Applicazioni N. 22, Accademia Nazionale dei Lincei, Roma (1976).

\bibitem{CKKY}
A.~Clark, A.~Karch, P.~Kovtun and D.~Yamada,
``Construction of bosonic string theory on infinitely curved anti-de  Sitter space,''
arXiv:hep-th/0304107.

\bibitem{LZ}
U.~Lindstrom and M.~Zabzine,
``Tensionless Strings, WZW Models at Critical Level and Massless Higher Spin
Fields,''
hep-th/0305098.

\bibitem{FL}
E.~S.~Fradkin and V.~Y.~Linetsky,
Phys.\ Lett.\ B {\bf 261} (1991) 26.

\bibitem{V} 
M.~A.~Vasiliev,
``Nonlinear equations for symmetric massless higher spin fields in  (A)dS(d),''
arXiv:hep-th/0304049.

\bibitem{HS}
M.~A.~Vasiliev,
arXiv:hep-th/9910096.
E.~S.~Fradkin and M.~A.~Vasiliev,
Nucl.\ Phys.\ B {\bf 291}, 141 (1987).
M.~A.~Vasiliev,
Nucl.\ Phys.\ B {\bf 616}, 106 (2001)
[Erratum-ibid.\ B {\bf 652}, 407 (2003)]
[arXiv:hep-th/0106200].
A.~K.~Bengtsson, I.~Bengtsson and L.~Brink,
Nucl.\ Phys.\ B {\bf 227}, 31 (1983).
V.~E.~Lopatin and M.~A.~Vasiliev,
Mod.\ Phys.\ Lett.\ A {\bf 3}, 257 (1988).
M.~A.~Vasiliev,
Annals Phys.\  {\bf 190}, 59 (1989).
M.~A.~Vasiliev,
Int.\ J.\ Mod.\ Phys.\ A {\bf 6}, 1115 (1991).
R.~R.~Metsaev,
Phys.\ Lett.\ B {\bf 354}, 78 (1995).
R.~R.~Metsaev,
Phys.\ Lett.\ B {\bf 419}, 49 (1998)
[arXiv:hep-th/9802097].
R.~R.~Metsaev,
Phys.\ Lett.\ B {\bf 531}, 152 (2002)
[arXiv:hep-th/0201226].
T.~Biswas and W.~Siegel,
JHEP {\bf 0207}, 005 (2002)
[arXiv:hep-th/0203115].

\bibitem{BMV}
L.~Brink, R.~R.~Metsaev and M.~A.~Vasiliev,
Nucl.\ Phys.\ B {\bf 586}, 183 (2000)
[arXiv:hep-th/0005136].

\bibitem{md}
M.~Plyushchay, D.~Sorokin and M.~Tsulaia,
JHEP {\bf 0304} (2003) 013
[arXiv:hep-th/0301067].

\bibitem{mm}
I.~L.~Buchbinder, A.~Pashnev and M.~Tsulaia,
Phys.\ Lett.\ B {\bf 523}, 338 (2001)
[arXiv:hep-th/0109067].
I.~L.~Buchbinder, A.~Pashnev and M.~Tsulaia,
arXiv:hep-th/0206026.

\bibitem{Gross2} 
D.~J.~Gross and P.~F.~Mende,
Phys.\ Lett.\ B {\bf 197} (1987) 129.
D.~J.~Gross and P.~F.~Mende,
Nucl.\ Phys.\ B {\bf 303} (1988) 407.

\bibitem{CW}
C.~G.~Callan and F.~Wilczek,
Nucl.\ Phys.\ B {\bf 340} (1990) 366.

\bibitem{Witten}
E.~Witten,
Nucl.\ Phys.\ B {\bf 268} (1986) 253.

\bibitem{GJ}
D.~J.~Gross and A.~Jevicki,
Nucl.\ Phys.\ B {\bf 283} (1987) 1.

\bibitem{holo}
I.~R.~Klebanov and A.~M.~Polyakov,
Phys.\ Lett.\ B {\bf 550} (2002) 213
[arXiv:hep-th/0210114].

\bibitem{adscft}
F.~Kristiansson and P.~Rajan,
JHEP {\bf 0304}, 009 (2003)
[arXiv:hep-th/0303202].
A.~Y.~Segal,
Nucl.\ Phys.\ B {\bf 664}, 59 (2003)
[arXiv:hep-th/0207212].
E.~Sezgin and P.~Sundell,
JHEP {\bf 0207}, 055 (2002)
[arXiv:hep-th/0205132].
E.~Sezgin and P.~Sundell,
Nucl.\ Phys.\ B {\bf 644}, 303 (2002)
[Erratum-ibid.\ B {\bf 660}, 403 (2003)]
[arXiv:hep-th/0205131].
S.~S.~Gubser, I.~R.~Klebanov and A.~M.~Polyakov,
Nucl.\ Phys.\ B {\bf 636}, 99 (2002)
[arXiv:hep-th/0204051].
E.~D'Hoker and D.~Z.~Freedman,
arXiv:hep-th/0201253.
A.~A.~Tseytlin,
Theor.\ Math.\ Phys.\  {\bf 133}, 1376 (2002)
[Teor.\ Mat.\ Fiz.\  {\bf 133}, 69 (2002)]
[arXiv:hep-th/0201112].
A.~Mikhailov,
arXiv:hep-th/0201019.
E.~Sezgin and P.~Sundell,
JHEP {\bf 0109}, 025 (2001)
[arXiv:hep-th/0107186].
E.~Sezgin and P.~Sundell,
JHEP {\bf 0109}, 036 (2001)
[arXiv:hep-th/0105001].
B.~Sundborg,
Nucl.\ Phys.\ Proc.\ Suppl.\  {\bf 102}, 113 (2001)
[arXiv:hep-th/0103247].
D.~A.~Lowe,
JHEP {\bf 0004}, 011 (2000)
[arXiv:hep-th/0003042].

\bibitem{also}
M.~Bianchi, J.~F.~Morales and H.~Samtleben,
JHEP {\bf 0307}, 062 (2003)
[arXiv:hep-th/0305052].
E.~Sezgin and P.~Sundell,
arXiv:hep-th/0305040.
R.~G.~Leigh and A.~C.~Petkou,
JHEP {\bf 0306}, 011 (2003)
[arXiv:hep-th/0304217].
N.~V.~Suryanarayana,
JHEP {\bf 0306}, 036 (2003)
[arXiv:hep-th/0304208].
S.~R.~Das and A.~Jevicki,
Phys.\ Rev.\ D {\bf 68}, 044011 (2003)
[arXiv:hep-th/0304093].
A.~C.~Petkou,
JHEP {\bf 0303}, 049 (2003)
[arXiv:hep-th/0302063].
D.~Birmingham, I.~Sachs and S.~N.~Solodukhin,
Phys.\ Rev.\ D {\bf 67}, 104026 (2003)
[arXiv:hep-th/0212308].
L.~Girardello, M.~Porrati and A.~Zaffaroni,
Phys.\ Lett.\ B {\bf 561}, 289 (2003)
[arXiv:hep-th/0212181].
S.~S.~Gubser and I.~R.~Klebanov,
Nucl.\ Phys.\ B {\bf 656}, 23 (2003)
[arXiv:hep-th/0212138].
J.~Engquist, E.~Sezgin and P.~Sundell,
Nucl.\ Phys.\ B {\bf 664}, 439 (2003)
[arXiv:hep-th/0211113].
T.~Leonhardt, A.~Meziane and W.~Ruhl,
Phys.\ Lett.\ B {\bf 555}, 271 (2003)
[arXiv:hep-th/0211092].

\endbib
\end{document}